\documentclass{article}

\usepackage[dvips]{graphicx}
\expandafter\ifx\csname package@font\endcsname\relax\else
 \expandafter\expandafter
 \expandafter\usepackage
 \expandafter\expandafter
 \expandafter{\csname package@font\endcsname}%
\fi
\begin{document}

\title{Earthquakes temporal occurrence: a statistical study}

\maketitle

\author{{\bf Cataldo Godano}}

\author{Department of Environmental Sciences, Seconda Universita' di
Napoli Caserta,

Italy

INFM UdR of Naples and Coordinate Group SUN}

\author{{\bf Lucilla de Arcangelis}

Department of Information Engineering, Seconda Universita' di
Napoli Aversa,

Italy

INFM UdR of Naples and Coordinate Group SUN}

\begin{abstract}

The distribution of inter-occurrence time between seismic events
is a quantity of great interest in seismic risk assessment. We
evaluate this distribution for different models of earthquakes
occurrence and follow two distinct approaches: The non homogeneous
Poissonian and the non Poissonian one. In all cases we obtain
either a power law or a power law damped by an exponential factor
behaviour. This feature of the distribution makes impossible any
prediction of earthquakes occurrence. Nevertheless it suggests the
interpretation of the earthquake occurrence phenomenon as due to
some non-linear dynamics to be further investigated.
\end{abstract}

\section{Introduction}

Seismology can be defined as the science of earthquakes and
studies mainly the physics of the earthquakes sources, the elastic
wave propagation and the occurrence of earthquakes in space, time
and energy. The investigation of earthquakes sources and wave
propagation are based on the analysis of the seismograms under the
assumption that linear theory of elasticity holds and are treated
in a huge amount of literature (see e.g. Ref. \cite{lay} and
references therein). On the other hand the study of earthquake
occurrence regards the construction and the investigation of
seismic catalogues, whose content is generally the time of
occurrence, the location and the energy of earthquakes. The great
interest dedicated by many researchers to the statistics of these
quantities is obviously linked to the interest in predicting the
time, the location and the energy of the next earthquake. These
questions are still rather unclear and we shall discuss some
aspects at the origin of the question. In particular, we suggest
that difficulties in prediction are intrinsic to the occurrence
mechanism.

The energy release in a seismic event is generally expressed by
the magnitude, which is proportional to the logarithm of the
maximum amplitude of the recorded seismic signal. The distribution
of magnitudes is described by an exponential law usually referred
as the Gutenberg-Richter distribution, by the names of the
researchers who firstly observed this feature of seismic
catalogues. They found that the frequency of occurrence of
earthquakes with magnitude greater then $M$ behaves as

\begin{equation}
{\rm log} N=a-bM
\end{equation}

where $a$ indicates the overall seismicity and $b$ is a scaling
parameter which has typically values close to one. Fluctuations
(up to 30\%) of the value of $b$ around its typical value are
widely observed depending on the catalogue, the estimating method
and the magnitude range \cite{FrolDav}. Regional variation of the
$b$ value are also observed by many authors (see e. g.
\cite{Hattori,Kronrod}). Such features could be an indication that
scaling properties of earthquakes are not universal.

However, it has been suggested that this discrepancy in $b$ values
could be due to systematic errors in magnitude determination
\cite{Utsu,Kagan}. For this reason, it is often studied the
distribution of the seismic moment $M_{0}$ defined as $M_0=\mu
A\Delta u$ where $\mu$ is the shear modulus, $A$ is the area of
the seismic fault involved and $\Delta u$ is the slip of the fault
due to a single seismic event
\cite{Kagan,Pacheco,FrolDav,Okal,Sornette1}. This is a more
physical quantity compared to the magnitude and can be obtained by
inverting the seismic signals with a standard procedure
\cite{Dzievonski}. The cumulative distribution of the seismic
moment exhibit a power law behavior followed by a sharp cutoff
after a $M_0^{max}$ well represented by a Gamma distribution

\begin{equation}
f(M_{0})=CM_{0}^{-(1+\gamma)}e^{-M_0\over{M_0^{max}}}
\end{equation}

where $C$ is a constant and $\gamma$ is a scaling parameter,
having a very stable value $\gamma \simeq 0.7$
\cite{Kagan,GodPin}.

The fractal nature of the spatial distribution of earthquakes has
been shown for the CMT catalogue \cite{KagKno,Sadovskiy} and the
fractal dimension $D_f$ of the hypocenter distribution has been
determined. However, the fractal dimension could be not sufficient
to describe all the scaling properties in a given problem and a
spectrum of dimensions turns out to be necessary in order to fully
characterize the scaling behavior. If this happens, the
distribution is said to be multifractal \cite{Paladin}.It has been
shown \cite{Hirata,Hirabayashi} that the spatial distributions of
earthquakes in Kanto region, in eastern Japan, in California and
in Greece have a multifractal structure. The multifractal
distribution of earthquakes hypocenters has been also confirmed
for some Italian regions \cite{GodBot,GodVil} and it has been
suggested \cite{Rossi,De Rubeis} that the temporal changes of
$D_f$ could be a good precursor parameter for earthquake
occurrence prediction.

On the other hand, the rate of occurrence of seismic events in
time has been widely investigated (see e.g.
\cite{Knopoff,Singh,Kuo}) because the existence of a
characteristic time scale could make possible the prediction of
the next earthquake. Unfortunately this is not the case since the
distribution of waiting times between earthquakes exhibits a two
power law behavior \cite{Udias,Smalley,ebel}(figure 1). More
precisely, the data exhibit a first power law regime with an
exponent close to 0.8 for both catalogues, followed by a second
power law with an exponent 1.8 for Colfiorito and 2.6 for
California. The experimental results suggest that inter-arrival
times are possible at all time scales, making the prediction of
earthquakes occurrence extremely difficult. The non-Poissonian
behavior of earthquake occurrence is clearly due to the phenomenon
of clustering, suggested by the power law distribution of the
inter-arrival times. In fact, for a Poissonian process, the
distribution would have an exponential behavior.

\begin{figure}
\caption{The experimental distribution of the waiting times for a)
the Californian catalogue of earthquakes (De Natale et al., 2003),
b)the Colfiorito (Italy) sequence(De Natale et al.,
2003).\label{figure:figura1}}
\end{figure}

The existence of power law behaviors and the multifractal
distribution of hypocenters, leads to the interpretation of
earthquake as a critical phenomenon \cite{Sornette, Bak, main},
proposing a new point of view for the features of earthquake
occurrence.

Finally in recent years it has been proposed that Self Organized
Criticality (SOC) \cite{Bak1} could explain the observed self -
similar properties of earthquakes \cite{Sornette, Bak}, in
particular could be able to reproduce the earthquake size
distribution. Within this approach, the lithosphere structure
derives from the self-organization of the earth crust in a
continental plate. A field theory can be also derived from
symmetry and conservation laws to explain the earthquakes size
distribution and viewing the earthquakes as fluctuation of the
elastic energy in the crust \cite{Sornette2}.

In this paper we focus on the waiting time distribution, we derive
the analytical expression of the inter-arrival time distribution
for some existing models and we discuss the scaling behavior of
the distributions depending on parameters. More precisely, in
Section 2 we shortly present some of the most commonly used models
and, in section 3, we derive the analytical expression of the
probability density function using a non homogeneous Poissonian
approach for the Omori law and the ETAS (Epidemic Type Aftershock
Sequences) model. In section 4 we derive the p.d.f. using a non
Poissonian approach for the Omori law and the Poisson Generalized
model. In all the cases there exists a choice of the parameters
fitting some data set. In the final section we present conclusions
and considerations for earthquakes occurrence prediction.

\section{The earthquake clustering and the Omori law}

It is widely observed that earthquakes tend to occur in bursts.
These bursts may start suddenly immediately following a large main
event, giving rise to the well known main - aftershocks sequences,
or may build up and die very gradually in time, generating swarms
of events. The most important implication of this kind of
occurrence is that we cannot assume a Poissonian occurrence of
earthquakes, where a Poissonian process is characterized by a
constant rate of occurrence, but rather a clustered one. In his
pioneering paper, Omori \cite{omori} investigated the problem of
earthquake occurrence within a single cluster of events and
proposed that the non-Poissonian behavior of seismic catalogues
could be well fitted with the Omori law, stating that the number
of aftershocks $n(t)$ decays in time as

\begin{equation}
n(t)={k\over {{(t+c)}^p}}
\end{equation}

where $p$ is generally very close to 1 ranging from $0.7$ to
$1.7$, $c$ is an initial time introduced in order to avoid the
divergence at $t = 0$ and $k=n(0)c^p$ is an experimental constant.

A widely used approach to earthquakes clustering is provided by
"trigger model" \cite{Vere}. This assumes a Poissonian occurrence
of main events, whereas the occurrence of the "triggered"
earthquakes is described in terms of a correlation function
$g(t-t_i)$, where $t_i$ is the time of occurrence of the $i$-th
event. The function $g(t-t_i)$ describes the correlation of each
event occurring at time $t$ with all the events occurred at
previous times. Thus the rate of occurrence will be

\begin{equation}
\lambda=\mu + \sum_{i:t_i<t}{g(t-t_i)}
\end{equation}

where $\mu$ is the Poissonian rate of the main events. Among the
trigger models a widely used one is the Poisson Generalised model
\cite{Toksoz}: this assumes the sequence of events as composed by
uncorrelated main events which generate clusters of aftershocks
distributed as the Pareto power law \cite{Vere}

\begin{equation}
q(j)= {{j^{-\beta}}\over {\zeta(\beta)}}
\end{equation}

where $\zeta(\beta)$ is the Riemann function and $j$ is the number
of events in the cluster. This approach has been applied for many
areas of the world, as California \cite{Toksoz}, Messina Strait
area \cite{Bottari} and Campi Flegrei (Italy) \cite{Zollo}, in
order to determine the $\beta$ value, found to be between 2.5 and
4.

A more appropriate choice of $g(t-t_i)$ is provided by the ETAS
model \cite{ogata}, which considers the existence of many clusters
described by the Omori law. The model states that the intensity
function (the rate of occurrence) of the earthquakes is given by:

\begin{equation}
\lambda=\mu+\sum_{i:t_i<t}{{ke^{\alpha(M_i-M_0)}}\over{(t-t_i+c)^p}}
\end{equation}

where $\mu$ is again the Poissonian rate , $\alpha$ an
experimental constant, $M_0$ is the smallest magnitude in the
catalogue and $M_i$ is the magnitude of the i-th event. The
meaning of equation (6) is that each earthquake can generate "its
own aftershocks" and that the number of these aftershocks depends
exponentially on the magnitude of the "main". In other words the
clustering degree varies in time, leading to a clustering within
the clustering.

A completely different approach is the fractal one \cite{Smalley}.
A Poissonian process would fill stochastically all the temporal
axis and thus would have a fractal dimension equal to 1, whereas a
clustered process is characterized by a fractal dimension less
then one depending on the clustering degree. Using the box
counting method it has been found that the New Hebrides seismicity
is clustered with a fractal dimension ranging between 0.126 and
0.255\cite{Smalley}. Moreover other authors \cite{Godano} found
that many catalogues in the world have a multifractal distribution
of inter-arrival times. This result is in good agreement with the
predictions of the ETAS model.

\section{The non homogeneous Poissonian approach}

The problem of earthquakes inter-arrival time distribution has
never been treated from the theoretical point of view. In order to
explain the temporal clustering properties of seismic events
occurrence, the most of efforts were dedicated to the study of the
rate of occurrence. Nevertheless the waiting time distribution is
very important in the seismic risk assessment because it is very
useful in the definition of the probability of the occurrence of
next earthquake. In this section we derive the distribution of the
waiting times for a single cluster following the Omori law and for
the ETAS model.

The cumulative distribution of waiting times $F(\Delta t)$ can be
written as \cite{Cox}

\begin{equation}
F(\Delta t)=1-F_0(\Delta t)
\end{equation}

where $F_0(\Delta t)$ is the probability of observing zero events
in $\Delta t$. Since for a Poissonian process

\begin{equation}
F_0(\Delta t)=e^{-\mu \Delta t}
\end{equation}

the probability density function (p.d.f.) is found to be

\begin{equation}
f(\Delta t)={{dF(\Delta t)}\over {d\Delta t}}=\mu e^{-\mu \Delta
t}
\end{equation}

which is the well known result for a Poissonian process.

This approach can be generalized also for processes for which
$\mu$ is not constant in time and we shall have a non homogeneous
Poissonian process. In this case the probability of having zero
events in $\Delta t$ is given by

\begin{equation}
F_0(\Delta t)=e^{-\int_0^{\Delta t}{\lambda(t) dt}}
\end{equation}

where $\lambda (t)$ is the time dependent rate of occurrence thus
inserting the (10) into the (7) and the result into the (9) we
obtain the waiting times p.d.f. Note that the Poissonian behavior
is more restrictive then the independence of events, since it is
obtained under the assumption that the probability of observing
more than one event in any small time interval, is negligible.

As a first application of this approach we shall derive the
waiting time distribution within a cluster of events. In this case
the rate of occurrence $\lambda (t)$ is given by the Omori law,
thus for $p\neq1$

\begin{equation}
F_0(\Delta t)=e^{-\int_0^{\Delta t}{{{k}\over{(t+c)^p}} dt}}
\end{equation}

and the p.d.f. of the ${\Delta t}$ will be

\begin{equation}
f(\Delta t)=ke^{{kc^{1-p} \over 1-p}}{(\Delta
t+c)}^{-p}e^{{-{k}\over{1-p}}(\Delta t+c)^{1-p}}
\end{equation}

which, except for some constant factors, is a Weibull
distribution, i.e. a power law damped by a stretched exponential
decay. Figure 2 shows the p.d.f. for different parameter value:
the Omori law exponent $p$ controls the decaying exponential
factor which is dominant at long times for ${(\Delta
t+c)^{1-p}}>{1-p \over k}$ when $p<1$ and at short times for
${(\Delta t+c)^{1-p}}<{{1-p}\over{k}}$ when $p>1$. We observe that
the p.d.f. for $p>1$ are not in agreement with experimental
observations.

\begin{figure}
\caption{The p.d.f. of the waiting time within a cluster of events
for different values of the model parameter\label{figure:figura1}}
\end{figure}

In the case $p=1$ from equation (11) we obtain

\begin{equation}
F_0(\Delta t)=\Big({\Delta t+c \over c}\Big)^{-k}
\end{equation}

and

\begin{equation}
f(\Delta t)=kc^k\Big({\Delta t+c \over c}\Big)^{-k-1}
\end{equation}

which is a power law and does not exhibit any exponential decay as equation (12).\\

A more complex formula is obtained if we adopt the ETAS model. In
this case we consider the existence of many clusters of events as
described in section 2. The rate of occurrence is given by the
equation (6). If we take the continuum limit, that is

\begin{equation}
\sum_{i:t_i<t}{{k(M)}\over{(t-t_i+c)^p}}\rightarrow\int_0^{t-\Delta
t}{{k(M)\over{(t-\tau+c)^p}}d\tau}
\end{equation}

where $k(M)=ke^\alpha (M_i-M_0)$ we will get for $p\neq1$

\begin{equation}
F_0(\Delta t)=e^{-\mu\Delta t+{k(M) \over(1-p)(2-p)}\{(\Delta
t+c)^{1-p}[\Delta t (1-p)+c]-c^{2-p}\} }
\end{equation}

which provides for the p.d.f.

$$f(\Delta t)=[\mu-{pk(M) \over 1-p}(\Delta t+c)^{-p}(\Delta t+{c
\over p})]$$
\begin{equation}
e^{-\mu\Delta t+{k(M) \over(1-p)(2-p)}\{(\Delta t+c)^{1-p}[\Delta
t (1-p)+c]-c^{2-p}\} }
\end{equation}

Equation (17) is well defined i.e. is a positive quantity, only
for $p>1$ and assumes the shape of a Weibull distribution (figure
3).

\begin{figure}
\caption{The p.d.f. of the waiting times for the ETAS model and
$p\neq 1$.\label{figure:figura2}}
\end{figure}

Analogously the case $p=1$ gives

\begin{equation}
F_0(\Delta t)=c^{ck}(\Delta t+c)^{-ck}e^{-\Delta t(\mu+k\rm
ln{{c+\Delta t}\over{4\Delta t}})}
\end{equation}

and for the p.d.f.

\begin{equation}
f(\Delta t)=\Big[\mu+k\rm ln{{c+\Delta t}\over{4\Delta
t}}\Big]c^{ck}(\Delta t+c)^{-ck}e^{-\Delta t(\mu+k\rm ln{{c+\Delta
t}\over{4\Delta t}})}
\end{equation}

We find again a power law damped by an exponential factor. Note
that the term in the square brackets is negative for $\mu<k\rm
ln{{c+\Delta t}\over{4\Delta t}}$ because $c+\Delta t<4\Delta t$,
however it is possible to obtain positive values of the p.d.f. for
$k<0.36$ if we set $\mu=0.5$ (figure 4).

It is noteworthy that the non homogeneous Poissonian approach does
not provide a good agreement with experimental data since does not
predict the two power regime shown in figure 1. This feature could
be due to the Poissonian assumption which assumes negligibly small
the probability of two events occurring in any small time
interval.

\begin{figure}
\caption{The p.d.f. of the waiting times for the ETAS model in the
case $p=1$ for different values of $k$\label{figure:figura3}}
\end{figure}

\section{The non Poissonian approach}

In this Section we derive the analytical expression of the p.d.f.
assuming only that the probability of cluster occurrence is
independent on the probability of earthquake occurrence within a
cluster. If we call $Q_n(\Delta t)$ the probability of having $n$
events in a cluster and $P_N(\Delta t)$ the probability of having
$N$ clusters in $\Delta t$, we will have

$$F_0(\Delta t)=P_0(\Delta t)Q_n(\Delta t)+$$
\begin{equation}
P_0(\Delta t)[1-Q_n(\Delta t)]+P_N(\Delta t)[1-Q_n(\Delta t)]
\end{equation}

The three terms in equation (20) represent respectively the
probability of having zero clusters of $n$ events, zero clusters
of zero events and $N$ clusters of zero events. Firstly we
determine the p.d.f. of the $\Delta t$ within a single cluster. In
this case $P_N(\Delta t)=1$ and $P_0(\Delta t)=0$. The number of
events $j$ in a time interval $\tau$  for $p\neq 1$ will be given
by

\begin{equation}
j(\tau)=\int_0^{\tau}{{k \over(t+c)^p}dt}={k \over
(1-p)}[(\tau+c)^{1-p}+c^{1-p}]
\end{equation}

Assuming the power law distribution (5) for $j$, we have

\begin{equation}
Q_n(\Delta t)={1\over {\zeta(\beta)}}\sum_{j=1}^n j^{-\beta}
\end{equation}

Noticing that in the continuum limit $\sum_j\rightarrow\int dt$
and neglecting the quantity $c^{1-p}$, we have

$$Q_n(\Delta t)={1 \over \zeta(\beta)}\Big({k \over
1-p}\Big)^{-\beta}\int_{(1-p)^{{1 \over 1-p}+c}}^{\Delta
t}{(\tau+c)^{-\beta(1-p)}d\tau}=$$
\begin{equation}
{k^{-\beta}(1-p)^\beta \over
\zeta(\beta)\delta}[(\Delta t+c)^\delta-(1-p)^{{1 \over 1-p}}]
\end{equation}

where $\delta=1-\beta(1-p)$. Finally we obtain the p.d.f.

\begin{equation}
f(\Delta t)={1 \over \zeta(\beta)}\Big({k
\over1-p}\Big)^{-\beta}(\Delta t+c)^{-\beta(1-p)}
\end{equation}

which is a power law well defined only for $p<1$. This constraint
is due to the assumption that $c^{1-p}\ll(\tau+c)^{1-p}$, which
implies from (21) that, if $p>1$, the number of events $j$ would
became negative.
 In the case $p=1$
we obtain a p.d.f. whose behavior is inconsistent with the
experimental data and thus will not be reported here.

Next we apply the non Poissonian approach to the "trigger" model
which assumes a Poissonian occurrence of clusters and a power law
decrease of the number of events within the clusters (equation
(5)). Under these assumptions we have

$$F_0(\Delta t)={e^{\mu\Delta t} \over \zeta(\beta)}\Big[\sum_{j=1}^n
j^{-\beta}+\Big({\zeta}({\beta}) -\sum_{j=1}^n j^{-\beta}\Big)+$$
\begin{equation}
{({\mu \Delta t)}^N \over
N!}\Big({\zeta}({\beta})- \sum_{j=1}^n j^{-\beta} \Big)\Big]
\end{equation}

Observing that for a Poissonian process the total number of
clusters is $N={\mu}{\Delta}t$ and that $({\mu \Delta}t)!$ in the
continuum limit becomes ${\Gamma}(1+{\mu \Delta}t)$, we have

\begin{equation}
F\Big({\tau \over \mu }\Big)={e^ {-\tau }\over
\zeta(\beta)}\Big[1+{\tau ^\tau \over {\Gamma}(1+{\lambda}
{\Delta}t)} \Big({\zeta}({\beta})-\sum_{j=1}^n j^{-\beta}
\Big)\Big]
\end{equation}

where $\tau=\mu\Delta t$. Using equation (22) in order to evaluate
$Q_n(\Delta t)$ and neglecting again the quantity $c^{(1-p)}$ we
obtain

$$f\Big({\tau \over \mu}\Big)=e^{-\tau}\Big\{\mu+{\tau^\tau \over \delta
\Gamma(1+\tau)\zeta(\beta)}$$
\begin{equation}
\big[K+\mu (\delta\zeta(\beta)+a\phi \Big({\tau\over\mu}\Big))
\Big({\Gamma'(1+\tau)\over\Gamma(1+\tau)}-\rm ln
\tau\Big)\Big]\Big\}
\end{equation}

where $a={k \over (1-p)}$, ${\delta}=1-\beta(1-p)$, $K=a{\delta}$,
$b=\Big({1-p \over k}+c^{(1-p)} \Big)^{1 \over 1-p}$ and
$\phi(x)=(b+c)^{-\delta}-(c+x)^{\delta}$. Equation (27) is a very
complex function and does not allow any simple fit of experimental
data. Moreover the number of parameters involved in the function
is too high for a stable fit. However we notice that it is
possible to find some plausible relations among some of the
parameters. For instance, $\mu$ and $k$ can be related since they
are both rates of occurrence: the first one concerns the cluster
occurrence, whereas the second one states how many earthquakes
occur at the beginning of a given cluster. In order to evaluate
equation (27), we choose $k=20{\mu}$. Any other choice for $k$ and
$\mu$ does not influence the shape of equation (28), but only the
level of the seismicity, that is the total number of events.
Obviously the value of $\mu$ and $k$, representing the time scale
in the system, implies as a consequence the value of the constant
$c$ of the Omori law and therefore we choose $c=0.3/k$. This means
that we have three free parameters $\mu$, $\beta$ and $p$. By
varying these parameters, we obtain two possible behaviors: either
a two power law regime or a two power law regime damped by an
exponential decay at high $\Delta t$.

\begin{figure}
\caption{The p.d.f. of the waiting times for the Poisson
generalized model at fixed values of $\beta=3.5$ and $p=0.85$ and
for different values of $\mu$.\label{figure:figura4}}
\end{figure}

In figure 5 we show the p.d.f. with fixed $\beta$ and $p$ for
different values of $\mu$. At lower values of $\mu$, therefore for
clusters more sparse in time, we observe the two power law regimes
behavior, whereas for increasing $\mu$ we observe the onset of an
exponential cut-off at long waiting times. Note that the exponents
are in the range 0.2 - 0.5 for the first power law and 1.3 - 1.9
for the second one.

Figure 6 shows the behavior of p.d.f. at fixed $\beta$ and $\mu$
for a range of values of $p$. In this case we observe for
decreasing values of the Omori exponent $p$, i.e. for clusters
lasting a longer time, the onset of an exponential cut-off at long
waiting times after the two power law regime. In this case the
exponents vary between 0.4 and 0.7 for the first power law and
between 1.0 and 1.3 for the second one. Any variation of the
parameter values does not change substantially the behavior in
figures 5 and 6. On the contrary, we will see that the p.d.f.
function is more sensitive to combined variations of $\beta$ and
$p$.

\begin{figure}
\caption{The p.d.f. of the waiting times for the Poisson
generalized model at fixed values of $\beta=3.5$ and $\mu=0.01$
and for different values of $p$.\label{figure:figura5}}
\end{figure}

\begin{figure}
\caption{The p.d.f. of the waiting times for the Poisson
generalized model at fixed values of $p=0.75$ and $\mu=0.01$ and
for different values of $\beta$.\label{figure:figura2}}
\end{figure}

Figure 7 shows for $p=0.75$ the onset of an exponential cut-off at
long waiting times for high values of $\beta$ as observed in
figures 5 and 6 (slopes are in the ranges 0.5 - 0.7 and 1.6 -
1.4). On the other hand, for $p=0.95$ the two power law behavior
is substantially insensitive to $\beta$ variations (figure 8).
This suggests that scaling properties of equation (27) are
dominated mainly by $p$ than by $\beta$. In this case the power
law exponents are 0.4 and 1.0.

\begin{figure}
\caption{The p.d.f. of the waiting times for the Poisson
generalized model at fixed values of $p=0.95$ and $\mu=0.01$ and
for different values of $\beta$.\label{figure:figura2}}
\end{figure}

The two power law regime is widely observed for many catalogues in
the world (figure 1) and generally interpreted as due to catalogue
incompleteness. Within the Poissonian Generalised approach we find
that the two power law behavior is quite robust with respect to
parameter changes. Therefore we suggest that this feature is an
intrinsic property of earthquake occurrence related to the P. G.
model.

\section{Conclusions}

We evaluate the probability density function of the
inter-occurrence time between earthquakes following two different
approaches. We first assume a non homogeneous Poissonian behavior
and find for different models of earthquakes occurrence always a
single power law, eventually followed by an exponential decay.

Next we investigate a non Poissonian approach for different
models. The obtained p.d.f. has a power law behavior in the case
of a single cluster of events described by the Omori law. On the
contrary, in the case of the Poisson Generalised model the p.d.f.
exhibit a more complex behavior depending on parameters. For all
values of $p\neq 1$ we find consistently a two power law regime.
This situation, occurring for small $\mu$, corresponds to long
waiting times between clusters of seismic events, which is the
situation more frequently observed in  nature. Depending on
parameters, the value of the exponents are in agreement with the
experimental data.

Moreover, for high values of $\beta$, i.e. fast decay in the
number of events in a single cluster, and a high Poissonian rate
$\mu$ the two power laws are followed by an exponential decay.
This feature characterizes a weak clustering in the distribution
of events in time or a frequent cluster occurrence.

The two power law behavior is observed for many catalogues
relatives to different areas in the world. This feature, often
interpreted as a sign of the incompleteness of the catalogue, is
here obtained as a specific characteristics of the p.d.f. for the
Poisson Generalised model. Finally we notice that for all the
discussed approaches and model the power law behavior implies the
absence of a characteristic inter-occurrence time and therefore
impossibility of any prediction of earthquake occurrence.

This work is part of the project of the Regional Center of
Competence "Analysis and Monitoring of the Environmental Risk"
supported by The European Community on Provision 3.16.

\end{document}